\documentclass[unnumsec,contemporary,large]{oup-authoring-template}%

\graphicspath{{Fig/}}

\theoremstyle{thmstyleone}%
\theoremstyle{thmstyletwo}%

\theoremstyle{thmstylethree}%

\usepackage{booktabs}

\usepackage{natbib}
\setcitestyle{sort&compress}
\newif\ifshowsupplemental
\showsupplementaltrue  

\begin{document}

\journaltitle{Bioinformatics}
\DOI{DOI HERE}
\copyrightyear{2025}
\pubyear{2025}
\access{Advance Access Publication Date: Day Month Year}
\appnotes{Application Note}

\firstpage{1}


\title[MHASS]{MHASS: Microbiome HiFi Amplicon Sequencing Simulator}

\author[1,$\ast$]{Rye Howard-Stone}
\author[1]{Ion I. M\u{a}ndoiu\ORCID{0000-0002-4818-0237}}

\authormark{Howard-Stone and M\u{a}ndoiu}

\address[1]{\orgdiv{School of Computing}, \orgname{University of Connecticut}, \orgaddress{\street{371 Fairfield Way}, \postcode{06269}, \state{CT}, \country{USA}}}

\corresp[$\ast$]{Corresponding author. \href{email:rye.howard-stone@uconn.edu}{rye.howard-stone@uconn.edu}}

\received{Date}{0}{Year}
\revised{Date}{0}{Year}
\accepted{Date}{0}{Year}


\abstract{
\textbf{Summary:} Microbiome HiFi Amplicon Sequence Simulator (MHASS) creates realistic synthetic PacBio HiFi amplicon sequencing datasets for microbiome studies, by integrating genome-aware abundance modeling, realistic dual-barcoding strategies, and empirically derived pass-number distributions from actual sequencing runs. MHASS generates datasets tailored for rigorous benchmarking and validation of long-read microbiome analysis workflows, including ASV clustering and taxonomic assignment.\\
\textbf{Availability and Implementation:} Implemented in Python with automated dependency management, the source code for MHASS is freely available at \url{https://github.com/rhowardstone/MHASS} along with installation instructions.\\
\textbf{Contact:} \href{rye.howard-stone@uconn.edu}{rye.howard-stone@uconn.edu} or \href{ion.mandoiu@uconn.edu}{ion.mandoiu@uconn.edu}\\
\textbf{Supplementary information:} Supplementary data are available at \textit{Bioinformatics} online.}

\keywords{microbiome, simulation, PacBio HiFi reads, amplicon sequencing, benchmarking}

\maketitle

\section{Introduction}

Long-read sequencing, particularly PacBio's High-Fidelity Circular Consensus Sequencing (CCS) technology, has revolutionized microbiome studies by enabling the use of long amplicons for taxonomic analysis \citep{Gehrig2022,ccs}. The ability to sequence multi-kilobase amplicons, such as the Titan-1\texttrademark{} region spanning 16S-ITS-23S ($\sim$2.5kb), offers unprecedented taxonomic resolution compared to traditional short-read approaches \citep{Graf2021}. However, rigorous benchmarking of computational tools for long-read amplicon analysis remains limited due to the scarcity of realistic synthetic datasets.

Existing simulation tools often omit critical features necessary for realistic microbiome amplicon data generation. Some, such as metaSPARSim \citep{metasparsim} and miaSIM\citep{miasim} produce solely an ASV count matrix while others \citep{pbsim3} focus on sequence-level simulation, ignoring abundances. Some programs simulate both abundances and sequencing effects, producing full synthetic datasets, such as CAMISIM \citep{camisim}, but do not include CCS technology. Other commonly excluded features include copy number-aware abundances accounting for multiple rRNA operons per genome, proper dual-barcode handling as used in multiplexed sequencing, and accurate pass-number distributions that directly influence CCS read accuracy. Furthermore, most simulators fail to capture the complex error profiles of PacBio HiFi reads, particularly sequence-specific patterns such as increased indel rates in homopolymer regions. As far as the authors are aware, currently there exists no tool that generates realistic multi-sample synthetic datasets for CCS amplicon sequencing of custom targets. A comparison of these programs is given in Supplemental Table~\ref{tab:feature_comparison}.

MHASS addresses these limitations by providing a modular pipeline that combines established tools with novel approaches to generate highly realistic PacBio HiFi amplicon datasets. We leveraged our recent AmpliconHunter tool \citep{ampliconhunter} for genome-aware amplicon extraction to test MHASS, which uses metaSPARSim \citep{metasparsim} for realistic abundance modeling, PBSIM3 \citep{pbsim3} for accurate subread simulation, and the PacBio CCS tool for consensus generation.

\section{Methods}

\subsection{Pipeline Overview}

MHASS employs a five-step pipeline to generate realistic PacBio HiFi amplicon datasets:

\textbf{Genome-Aware Abundance Simulation.}
MHASS utilizes metaSPARSim to generate abundance matrices at the genome level by calculating the expected abundance of each of its multiple ASVs, based on the input genome molarity as supplied by the user. This approach captures the biological reality that bacterial genomes often contain multiple, non-identical copies of the rRNA operon. Users can select from uniform, lognormal, power-law, or empirical abundance distributions based on actual microbiome profiles. The variability of an ASVs counts are given as a function of their intensity, using a negative binomial model fit to the R1 preset provided by metaSPARSim. See Supplemental Figure \ref{fig:variability} for more information.

\textbf{Template Generation with Barcodes.}
Each sample receives uniquely barcoded templates following PacBio's dual-barcode strategy:
\begin{verbatim}
A + ForwardBarcode + ForwardPrimer + ASV + 
ReverseComplement(ReversePrimer) + 
ReverseComplement(ReverseBarcode) + A
\end{verbatim}
This structure accurately reflects the sequencing library preparation process, enabling realistic demultiplexing evaluations by allowing the simulation to include errors in barcodes.

\textbf{Subread Simulation.}
MHASS addresses a key limitation of PBSIM3 (its restriction to a single pass number per run) by implementing a novel approach based on empirical pass-number distributions. We analyzed pass-number distributions from publicly available PacBio HiFi datasets \citep{ubiome,phylotag} and identified that they follow neither normal nor uniform distributions, but rather exhibit complex patterns dependent on insert length and sequencing conditions. MHASS samples from either empirical distributions extracted from real data or fitted lognormal distributions, running PBSIM3 separately for each pass number and parallelizing across CPU cores. In our validation experiments, number of passes was fit from real data (see Figure \ref{fig:np_distribution_comparison} for details), and subread accuracy was empirically optimized for each dataset using KL divergence minimization (Supplementary Figure \ref{fig:subread}). For all three datasets we retained PBSIM3’s PacBio RSII default error-type difference ratio (6:55:39 for substitution:insertion:deletion), as no Sequel II-specific defaults are published.

\textbf{CCS Consensus.}
Simulated subreads undergo CCS processing using PacBio tools (v6.4.0+), yielding consensus sequences with realistic base quality profiles. The CCS algorithm is applied identically to both simulated and real data.

\textbf{Data Consolidation and Cleanup.}
Final reads are combined into a single FASTQ file with proper formatting and metadata. Read headers maintain traceability to original templates while conforming to standard PacBio formatting conventions for compatibility with downstream analyses.

\subsection{Validation Datasets}

We validated MHASS using three 
datasets with known ground truth: Zymo/Titan-1 (Titan-1 amplicon defined by forward primer 5$^\prime$-AGRRTTYGATYHTDGYTYAG-3$^\prime$ and reverse primer 5$^\prime$-YCNTTCCYTYDYRGTACT-3$^\prime$, D6300 mock community, 8 genomes with even abundances, 96 samples, 186{,}167 total reads), ATCC/16S (full 16S amplicon defined by primers 27F (5$^\prime$-AGRGTTYGATYMTGGCTCAG-3$^\prime$) and 1492R (5$^\prime$-RGYTACCTTGTTACGACTT-3$^\prime$), MSA-1003 mock community, 20 genomes with staggered abundances, 192 samples, 2{,}468{,}174 total reads), and Phylotag/16S (full 16S amplicon also defined by primers 27F and 1492R, custom mock community, 23 genomes with highly staggered abundances, 5 samples, 113{,}709 total reads). The 16S datasets are publicly available \citep{atcc,phylotag}; the Zymo/Titan-1 dataset was provided to the authors by Intus Biosciences. More information on these datasets is given in Supplemental Tables~\ref{tab:performance}-
\ref{tab:phylotag}. The amplicon multiplicity patterns ('Amplitypes' \citep{ampliconhunter}) for these mock communities are depicted in Supplementary Figures \ref{fig:zymo_d6300_titan_amplitypes}-
\ref{fig:phylotag_16s_amplitypes}.

\subsection{Evaluation Metrics}

We evaluated simulation realism through multiple approaches: (1) Edit Distance Distributions - comparing the distribution of minimum global edit distances between reads and reference sequences; (2) Abundance Correlation - Pearson correlation between expected and observed ASV abundances at both genome and ASV levels; (3) Error Profiles - analysis of substitution, insertion, and deletion patterns as a function of position in the amplicon.

\section{Results and Discussion}

\subsection{Edit Distance Distributions Match Real Data}

\begin{figure*}[htbp]
  \centering
  \includegraphics[width=0.82\textwidth]{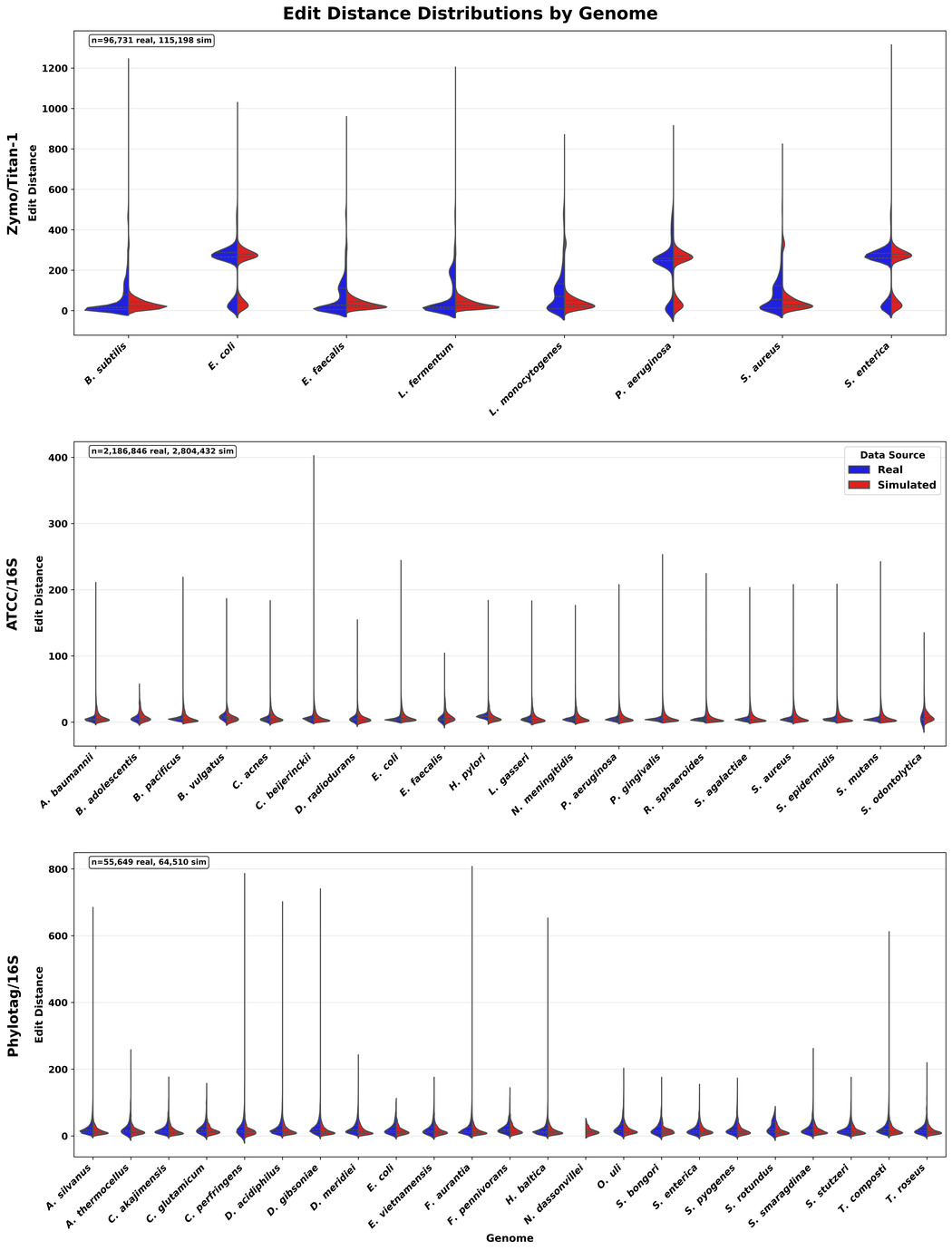}
  \caption{Edit distance distribution from read to nearest reference ASV by genome for both real and MHASS-simulated reads. Each panel corresponds to a different mock community: Zymo/Titan-1 (top), ATCC/16S (middle), and Phylotag/16S (bottom). Blue violins show distributions from real CCS data; red violins show MHASS-simulated reads. Total number of reads surviving QC steps for each dataset are indicated in the top left. No misassignment of simulated reads occurred; all simulated reads were found to be closest to the original ASV sequences from which they were simulated.}
  \label{fig:multi_panel_violins}
\end{figure*}

Across all three mock communities, MHASS generated reads with edit distance distributions that closely mirrored those observed in real sequencing data. Figure~\ref{fig:multi_panel_violins} presents per-genome violin plots of the edit distance between each read and its nearest reference ASV, with real data shown in blue and simulated data in red.

The Zymo/Titan-1 dataset displays a distinctive bimodal distribution, most pronounced in genomes like \textit{E. coli}, \textit{P. aeruginosa} and \textit{S. enterica}, where two distinct peaks emerge. MHASS successfully captured this characteristic pattern, indicating it replicates both low-error and high-error modes of CCS read distributions for complex amplicon mixtures.

In contrast, the ATCC/16S mock community exhibited unimodal distributions across genomes, reflecting the more conserved and less variable nature of the 16S amplicon sequences. MHASS closely reproduced the shape and spread of these distributions, with consistent error profiles observed across all 20 species.

Finally, the Phylotag/16S dataset posed the most stringent challenge due to its highly staggered taxonomic abundances. Nevertheless, MHASS maintained realistic error behavior even in low-abundance genomes. Although the rarest genome (\textit{N. dassonvillei}) was assigned no real reads, the overall shape and central tendency of most distributions remained well-aligned with real data.

\subsection{Abundance Correlation}

Expected and observed counts were correlated for both real and simulated data: Zymo/Titan-1 (even abundances): $R^2 = 0.662$ (real), $R^2 = 0.519$ (simulated); ATCC/16S (staggered): $R^2 = 0.956$ (real), $R^2 = 0.999$ (simulated); Phylotag/16S (highly staggered): $R^2 = 0.794$ (real), $R^2 = 0.944$ (simulated). Genome-level agreement between expected and observed abundances is visualized for both real and simulated data for all datasets in Supplementary Figure \ref{fig:abundance}. MHASS fits a negative binomial model to predict the squared coefficient of variation of each ASV between samples based on its expected abundance using the values in metaSPARSim's R1 preset \citep{metasparsim}, thus modeling realistic biological variability between replicates. Parameter fitting is shown in Figure \ref{fig:variability}. This choice provides moderate biological variability typical of microbiome data. The NA values frequently returned by metaSPARSim's parameter estimation procedure are emulated when the sampled variability from the negative binomial model would be less than 0. When variability is NA, metaSPARSim converts it to 0, resulting in no biological variability - all replicates receive identical abundance values before the technical sampling step.

\subsection{Accurate Modeling of Error Types}

Analysis of error types revealed that the secondary peak in Zymo/Titan-1 edit distances resulted from insertions concentrated at the ends of the reads (positions $\sim$2100--2500 of the alignment). Both real and simulated data showed elevated insertion rates in this region, with the proportion of insertions increasing from baseline to $>$90\% at position $\sim$2100 until the end of the alignment (see Supplemental Figure \ref{fig:errors_by_position}. This pattern was absent in 16S-only datasets, confirming its association with the Titan-1 amplicon.

\section{Implementation and Performance}

MHASS is implemented primarily in Python 3.6+ with the following key dependencies: metaSPARSim 1.1.2 \citep{metasparsim} for abundance simulation (R 4.0+), PBSIM3 \citep{pbsim3} for subread simulation, and PacBio CCS tools \citep{ccs} 6.4.0 for consensus calling. It also requires conda (23.3.1) and Samtools (1.10) for dependency management and sequence processing. 
Installation is automated through a bash script that creates appropriate conda environments and downloads required models, compiling all dependencies for the target system:
\begin{verbatim}
git clone https://github.com/rhowardstone/MHASS.git
cd MHASS
bash install_dependencies.sh
conda activate mhass
\end{verbatim}

\subsection{Performance Benchmarks}

Performance was measured on a virtual machine configured with 180 virtual cores and 374GB RAM running on a Dell PowerEdge R7525 server with two AMD EPYC 7552 48-core CPUs. Simulation time and maximum memory usage scale linearly with the number of simulated reads, with additional overhead from the empirical pass-number sampling approach: Zymo/Titan-1 dataset (96 samples, $\sim$2k reads/sample): 40 minutes; ATCC/16S dataset (192 samples, $\sim$13k reads/sample): 6.75 hours; Phylotag/16S dataset (5 samples, $\sim$23k reads/sample): 23 minutes. The runtime and memory usage statistics for each dataset are summarized in Supplementary Table \ref{tab:performance}. 

\subsection{Parameter Guidelines}

Key parameters affecting simulation realism include: \texttt{subread accuracy} (default 0.65, based on PBSIM3 calibration, see Figure \ref{fig:subread} for more information), \texttt{np distribution type} (`empirical' or `lognormal', \texttt{genome distribution} (choice significantly impacts abundance patterns), and \texttt{np min}, \texttt{np max} (pass number range, default: 2--59). Reproducible scripts to fit the parameters used in this evaluation are provided in our supplementary GitHub: \url{https://github.com/rhowardstone/MHASS_evaluation}.

\section{Conclusion}

MHASS addresses a critical need in the microbiome community for realistic long-read amplicon simulators. Our validation experiments demonstrate that MHASS accurately reproduces key characteristics of real PacBio HiFi data, including complex error patterns and abundance distributions. The presence of sequence-specific insertion patterns in Titan-1 amplicons highlights the importance of region-aware error modeling. This finding has implications for ASV calling algorithms, which may need to account for region-specific error rates to avoid oversplitting true biological variants.

Current limitations include: (1) PCR bias modeling remains simplistic, not accounting for primer-specific amplification efficiencies; (2) Chimera formation is not explicitly modeled; (3) Adapter dimer formation and other library preparation artifacts are not included. Future versions will address these limitations and extend support to other long-read platforms (ONT) and newer PacBio chemistries (Revio).

MHASS provides a much-needed tool for the microbiome research community, enabling rigorous benchmarking of analytical pipelines for PacBio HiFi amplicon sequencing. Its 
integration of genome-aware modeling, empirical realism, and streamlined usability makes it an essential resource for robust computational method development in microbiome research.

\bibliography{references}

\setcounter{figure}{0}
\setcounter{table}{0}
\renewcommand{\thefigure}{S\arabic{figure}}
\renewcommand{\thetable}{S\arabic{table}}

\refstepcounter{table}\label{tab:feature_comparison}
\refstepcounter{figure}\label{fig:variability}
\refstepcounter{figure}\label{fig:np_distribution_comparison}
\refstepcounter{figure}\label{fig:subread}
\refstepcounter{figure}\label{fig:zymo_d6300_titan_amplitypes}
\refstepcounter{figure}\label{fig:atcc_msa_1003_16s_amplitypes}
\refstepcounter{figure}\label{fig:phylotag_16s_amplitypes}
\refstepcounter{figure}\label{fig:abundance}
\refstepcounter{figure}\label{fig:errors_by_position}
\refstepcounter{table}\label{tab:performance}
\refstepcounter{table}\label{tab:zymo}
\refstepcounter{table}\label{tab:atcc}
\refstepcounter{table}\label{tab:phylotag}

\ifshowsupplemental
\clearpage
\onecolumn
\section*{Supplementary Figures}

\setcounter{figure}{0}
\setcounter{table}{0}
\renewcommand{\thefigure}{S\arabic{figure}}
\renewcommand{\thetable}{S\arabic{table}}

\begin{figure*}[htbp!]
    \centering
    \includegraphics[width=0.5\linewidth]{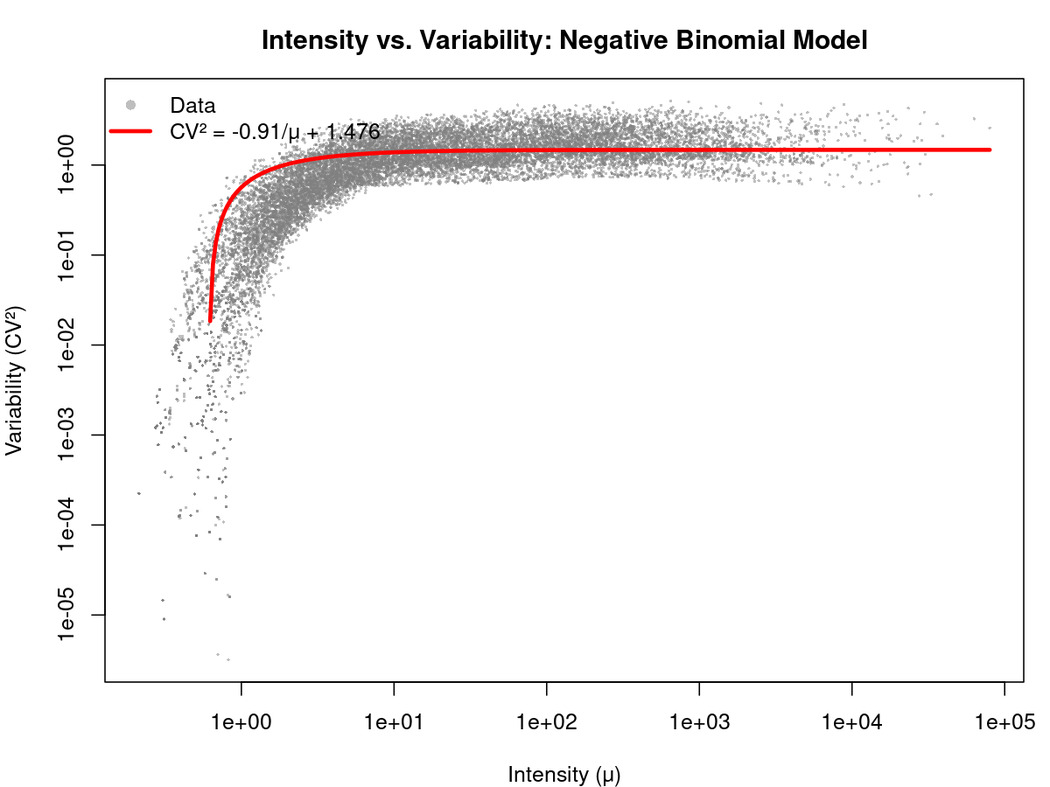}
    \caption{
    Variability model fitting for abundance simulation in MHASS. To determine appropriate variability parameters for metaSPARSim, we analyzed the relationship between intensity $\mu$ (mean abundance) and variability ($CV^2$) using the R1 dataset provided by metaSPARSim. The data exhibited a negative binomial-like relationship where $CV^2 = \beta_1/\mu + \beta_0$. Linear regression on $CV^2$ vs. $1/\mu$ yielded $CV^2 = -0.910/\mu + 1.476$ with $R^2=0.370$. As mean abundance increases, the coefficient of variation decreases, consistent with overdispersed count data. The negative coefficient for the $1/\mu$ term (-0.910) deviates from the theoretical value of 1.0 for a true negative binomial distribution, suggesting the microbiome abundance data follows a modified overdispersion pattern. Data points are shown in gray, the fitted regression line in red, and axes are log-transformed.
    }
    \label{fig:variability}
\end{figure*}

\begin{figure*}[htbp!]
    \centering
    \includegraphics[width=0.7\linewidth]{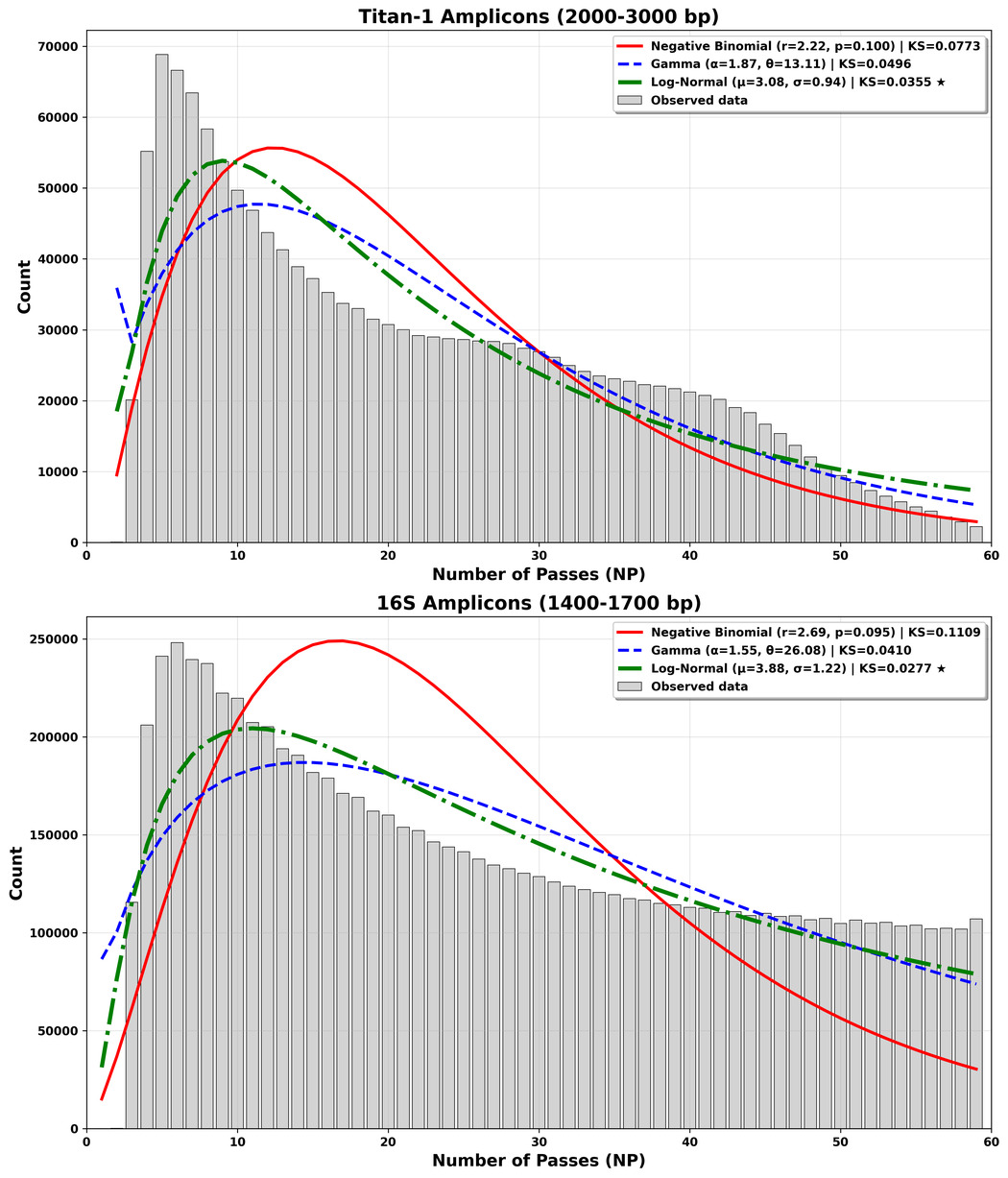}
    \caption{Number of passes from two datasets with CCS annotations are fit to three models. The top panel represents the distribution of number of passes for Titan-1 amplicons \citep{ubiome}, reads filtered to a length of 2,000-3,000 bp. The bottom panel represents the same distribution for a dataset of 16S amplicons \citep{kinnex}, filtered to 1400-1700 bp in length. Three models are fit to each distribution, with Log-Normal obtaining the lowest Kolmogorov–Smirnov statistic for both datasets.}
    \label{fig:np_distribution_comparison}
\end{figure*}

\begin{figure*}[htbp!]
    \centering
    \includegraphics[width=0.8\linewidth]{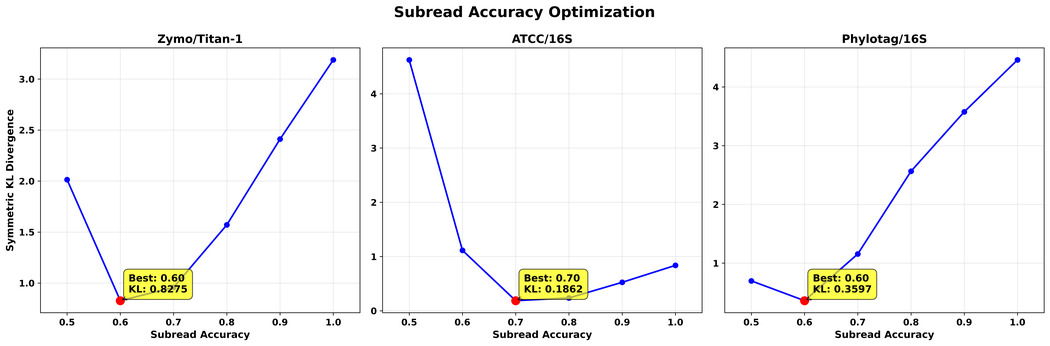}
    \caption{Subread accuracy optimization results for all three datasets (from left to right): Zymo/Titan-1, ATCC/16S, and Phylotag/16S. The subread accuracy parameter for PBSIM3 was optimized independently for each dataset by testing accuracy values from 0.5 to 1.0 in increments of 0.1, and running complete simulations for each value. The optimal accuracy was selected by minimizing the Kullback-Leibler (KL) divergence between the edit distance distributions of real and simulated reads. KL divergence between the distributions of minimum edit distances from reads to references for real vs. simulated reads are shown in blue. The optimal values are highlighted in red with and annotated with their KL divergence values and optimal subread accuracy: 0.6 for Zymo/Titan-1, 0.7 for ATCC/16S, and 0.6 for Phylotag/16S.}
    \label{fig:subread}
\end{figure*}

\begin{figure*}[htbp]
    \centering
    \includegraphics[width=0.6\linewidth]{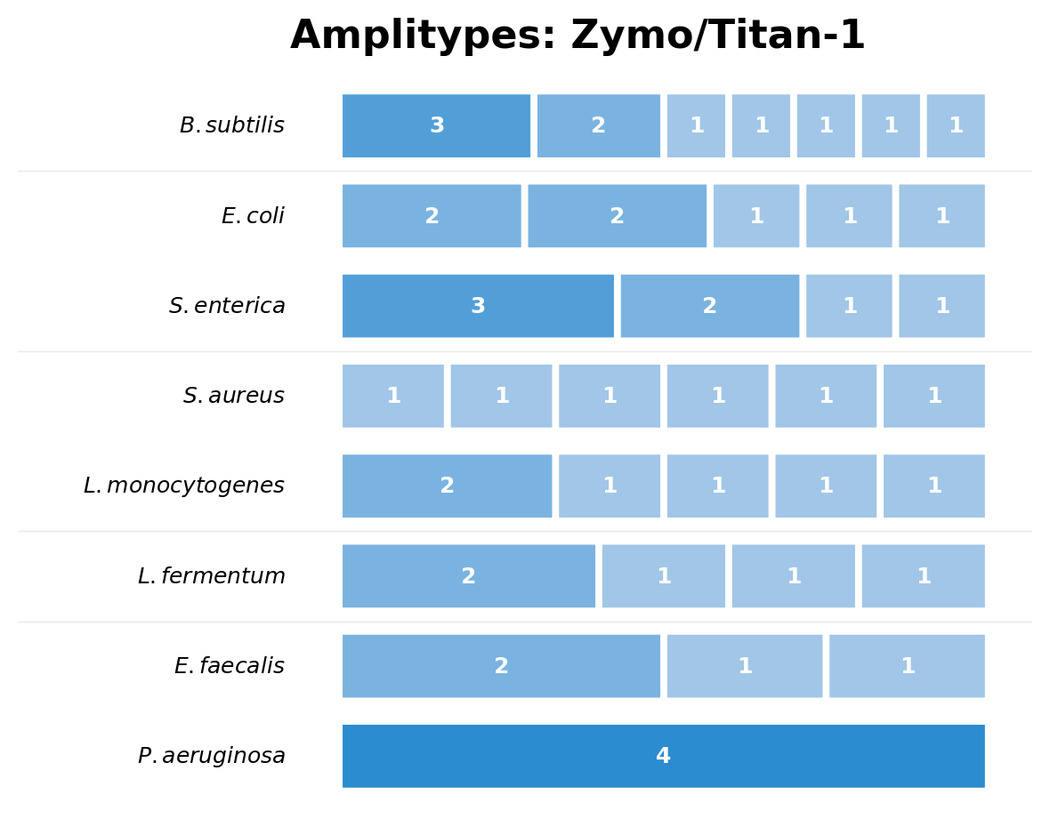}
     \caption{Amplicon multiplicity types detected in the Zymo/Titan-1 dataset, showing the multiple amplicon variants per genome with copy number.}
    \label{fig:zymo_d6300_titan_amplitypes}
\end{figure*}

\begin{figure*}[htbp]
    \centering
    \includegraphics[width=0.7\linewidth]{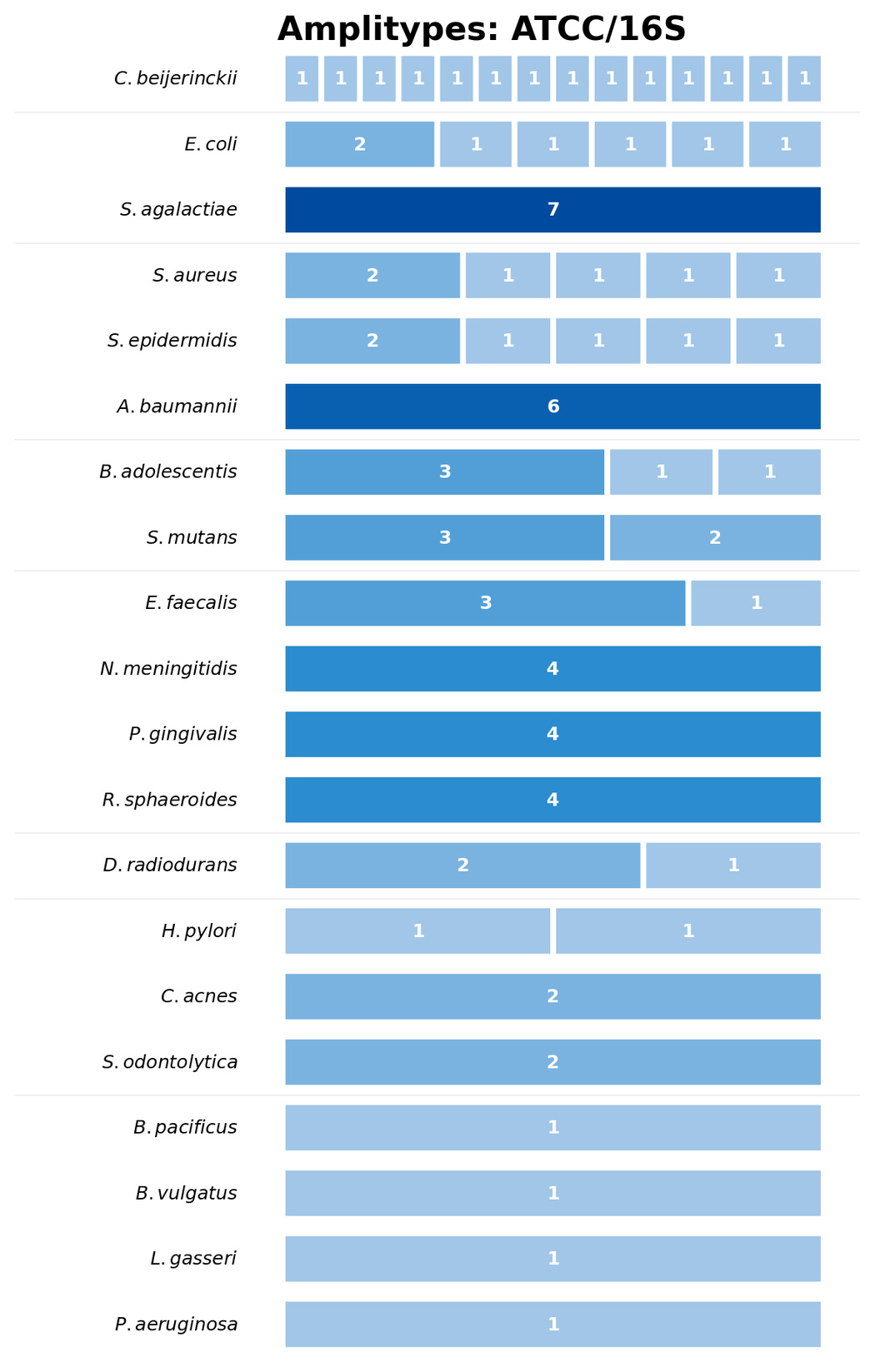}
    \caption{Amplicon multiplicity types detected in the ATCC/16S dataset, showing the multiple amplicon variants per genome with copy number.}
    \label{fig:atcc_msa_1003_16s_amplitypes}
\end{figure*}

\begin{figure*}[htbp]
    \centering
    \includegraphics[width=0.7\linewidth]{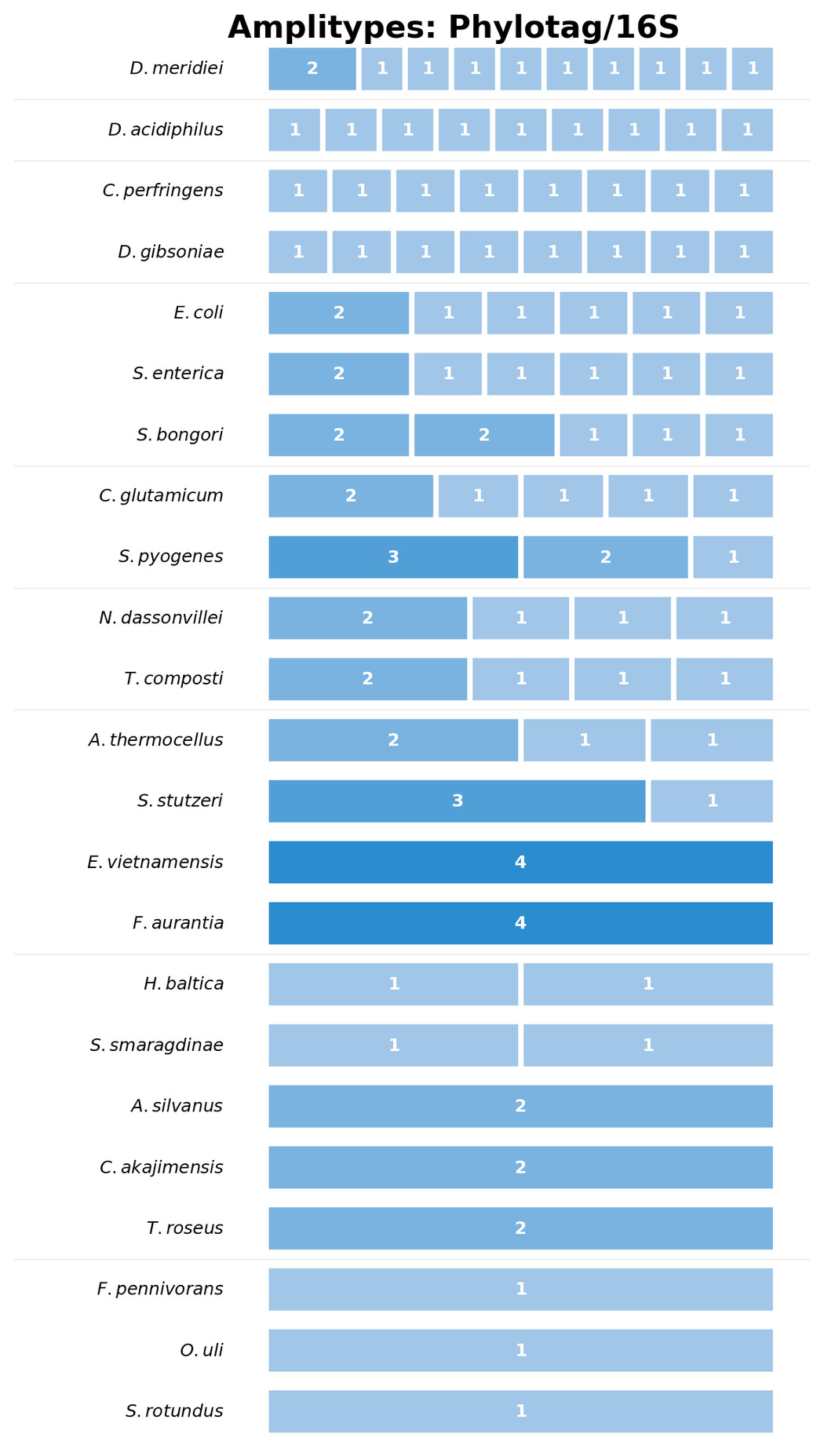}
    \caption{Amplicon multiplicity types detected in the Phylotag/16S dataset, showing the multiple amplicon variants per genome with copy number.}
    \label{fig:phylotag_16s_amplitypes}
\end{figure*}

\begin{figure*}[htbp]
    \centering
    \includegraphics[width=0.5\linewidth]{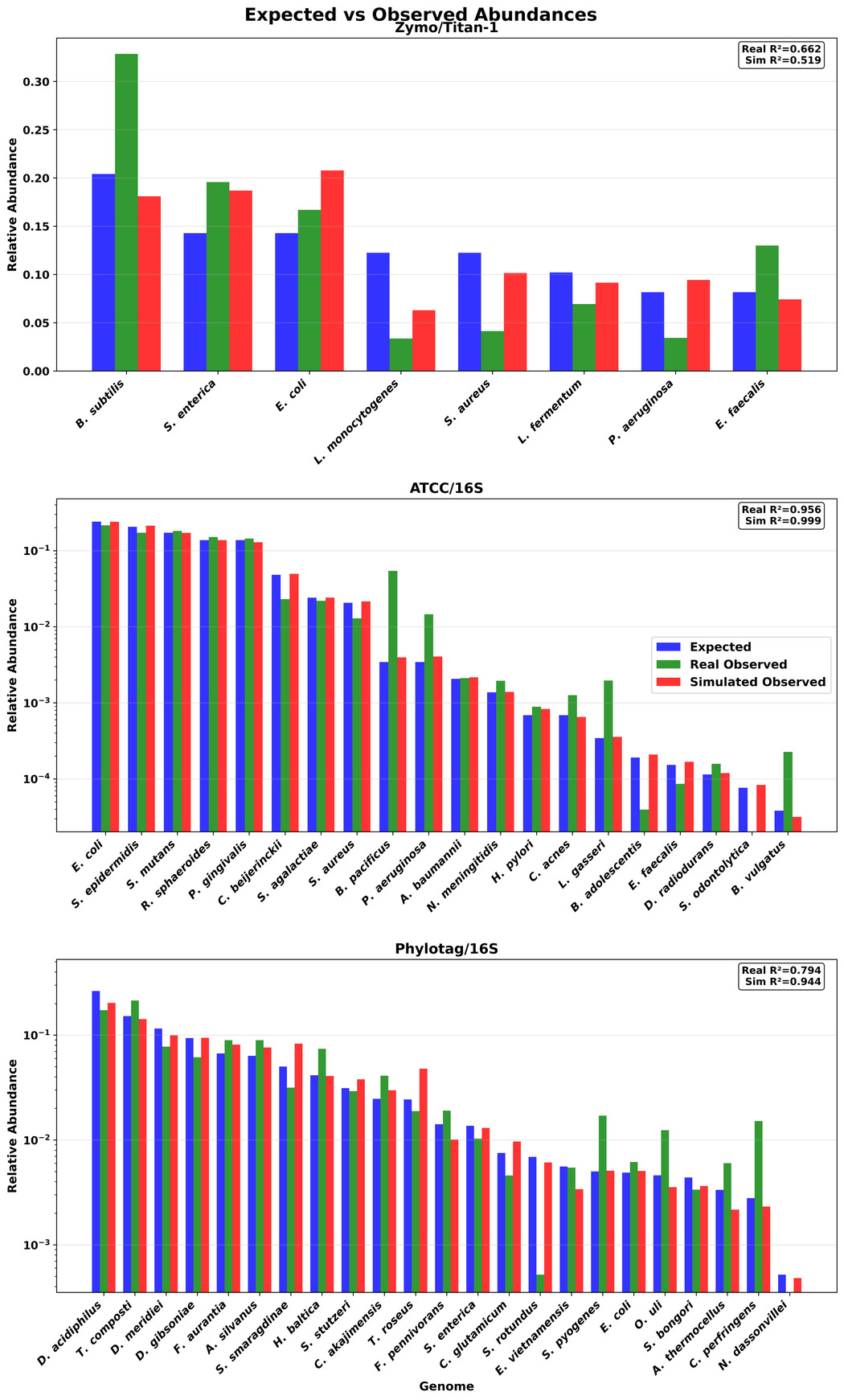}
    \caption{Bar plots showing expected (based on known mock community composition) versus observed ASV abundances at the genome level for both real and simulated data. Correlation coefficients ($R^2$) demonstrate high fidelity of abundance simulation across all three datasets. Abundances for the Zymo/Titan-1 and ATCC/16S datasets are known and published in their respective product sheets \citep{zymo,atcc}. The custom mock community, Phylotag/16S, was produced by the authors of \cite{phylotag}, who note pipetting errors in mixing. To circumvent this, we used their measurements of short read shotgun data as the source of truth for genome abundance, rather than molarity, as the authors suggest. Values were estimated from Supplementary Figure 4 of \cite{phylotag}, using \cite{graphreader}.}
    \label{fig:abundance}
\end{figure*}

\begin{figure*}[htbp]
    \centering
    \includegraphics[width=\linewidth]{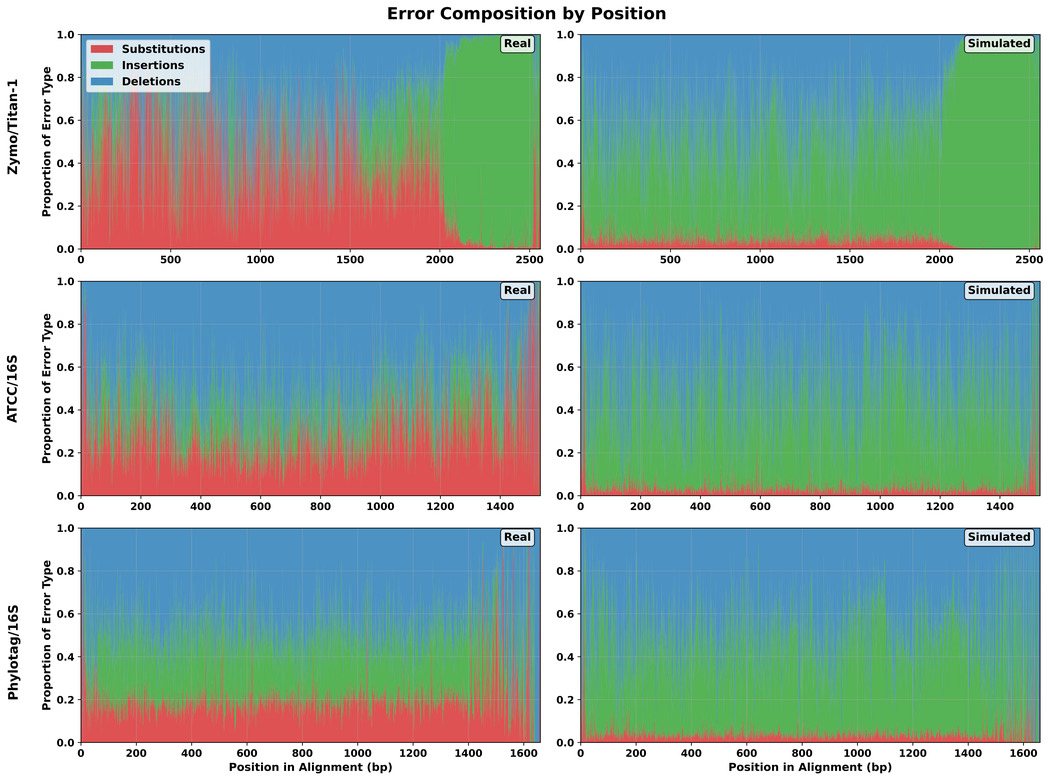}
    \caption{Each read is aligned globally to its nearest reference ASV. Proportion of errors by type and position in alignment is shown for all three datasets (from top to bottom): Zymo/Titan-1, ATCC/16S, and Phylotag/16S. Substitutions are shown in red, insertions in green, and deletions in blue. Real data is shown on the left, simulated data is shown on the right. A notable increase of insertions is seen in both real and simulated reads for the Zymo/Titan-1 dataset from approximately position 2100 to the end of the alignment. The proportion of substitutions in real data is much higher than would be implied by the default error ratio parameter supplied by PBSIM3 for the RSII \citep{pbsim3}.}
    \label{fig:errors_by_position}
\end{figure*}

\clearpage
\section*{Supplementary Tables}

\begin{table}[ht]
\centering
\caption{Feature comparison of microbiome simulation tools}
\begin{tabular}{lccccc}
\toprule
\textbf{Feature} & \textbf{metaSPARSim} & \textbf{PBSIM3} & \textbf{CAMISIM} & \textbf{miaSim} & \textbf{MHASS} \\
\midrule
Abundance                  & \checkmark &            & \checkmark & \checkmark & \checkmark \\
Copy \#                    &            &            &            &            & \checkmark \\
Barcoding                  &            &            &            &            & \checkmark \\
Sequencing errors          &            & \checkmark & \checkmark &            & \checkmark \\
CCS-compatible       & \checkmark & \checkmark &            & \checkmark & \checkmark \\
Pass number distribution   &            &            &            &            & \checkmark \\
\bottomrule
\end{tabular}
\label{tab:feature_comparison}
\end{table}

\begin{table}[ht]
\centering
\caption{Dataset information and computational resource usage for MHASS evaluation. 
Performance was measured on a 
virtual machine configured with 180 virtual cores and 374GB RAM running on a Dell PowerEdge R7525 server with two AMD EPYC 7552 48-core CPUs.
CPU usage indicates effective parallelization across multiple cores, with an average core usage of 100/180. Runtime includes all pipeline steps starting with ASV extraction from genomes, through to final FASTQ generation.}
\begin{tabular}{lccccc}
\toprule
Dataset & Platform & Total Reads & Samples & Runtime (h:mm:ss) & Peak Memory (GB) \\
\midrule
Zymo/Titan-1 & Sequel II & 186,167 & 96 & 0:40:23 & 1.8 \\
ATCC/16S & Sequel II & 2,468,174 & 192 & 6:45:55 & 8.8 \\
Phylotag/16S & RSII & 113,709 & 5 & 0:22:36 & 0.9 \\
\bottomrule
\end{tabular}
\label{tab:performance}
\end{table}

\begin{table}[ht]
\centering
\caption{Dataset Information - Zymo/Titan-1 (D6300)}
\begin{tabular}{lcc}
\toprule
\textbf{Species}        & \textbf{Abundance (\%)} & \textbf{Copy Number} \\
\midrule
Bacillus subtilis       & 12.5                    & 10                   \\
Enterococcus faecalis   & 12.5                    & 4                    \\
Escherichia coli        & 12.5                    & 7                    \\
Lactobacillus fermentum & 12.5                    & 5                    \\
Listeria monocytogenes  & 12.5                    & 6                    \\
Pseudomonas aeruginosa  & 12.5                    & 4                    \\
Salmonella enterica     & 12.5                    & 7                    \\
Staphylococcus aureus   & 12.5                    & 6                   
\end{tabular}
\label{tab:zymo}
\end{table}

\begin{table}[ht]
\centering
\caption{Dataset Information - ATCC/16S 
(MSA-1003)}
\begin{tabular}{llccc}
\toprule
\textbf{Species}                          & \textbf{Accession} & \textbf{Abundance (\%)} & \textbf{Copy Number} \\
\midrule
Acinetobacter baumannii (ATCC 17978)      & GCF\_902728005.1   & 0.18                    & 6                    \\
Bacillus pacificus (ATCC 10987)           & GCF\_031316815.1   & 1.8                     & 1                    \\
Bacteroides vulgatus (ATCC 8482)          & GCF\_028538915.1   & 0.02                    & 1                    \\
Bifidobacterium adolescentis (ATCC 15703) & GCF\_000010425.1   & 0.02                    & 5                    \\
Clostridium beijerinckii (ATCC 35702)     & GCF\_000767745.1   & 1.8                     & 14                   \\
Cutibacterium acnes (ATCC 11828)          & GCF\_000231215.1   & 0.18                    & 2                    \\
Deinococcus radiodurans (ATCC BAA-816)    & GCF\_000008565.1   & 0.02                    & 3                    \\
Enterococcus faecalis (ATCC 47077)        & GCF\_004006275.1   & 0.02                    & 4                    \\
Escherichia coli (ATCC 700926)            & GCF\_000364365.1   & 18                      & 7                    \\
Helicobacter pylori (ATCC 700392)         & GCF\_000008525.1   & 0.18                    & 2                    \\
Lactobacillus gasseri (ATCC 33323)        & GCF\_008868295.1   & 0.18                    & 1                    \\
Neisseria meningitidis (ATCC BAA-335)     & GCF\_000008805.1   & 0.18                    & 4                    \\
Porphyromonas gingivalis (ATCC 33277)     & GCF\_002892575.1   & 18                      & 4                    \\
Pseudomonas aeruginosa (ATCC 9027)        & GCF\_001294675.1   & 1.8                     & 1                    \\
Rhodobacter sphaeroides (ATCC 17029)      & GCF\_000015985.1   & 18                      & 4                    \\
Schaalia odontolytica (ATCC 17982)        & GCF\_000154225.1   & 0.02                    & 2                    \\
Staphylococcus aureus (ATCC BAA-1556)     & GCF\_032809245.1   & 1.8                     & 6                    \\
Staphylococcus epidermidis (ATCC 12228)   & GCF\_022869565.1   & 18                      & 6                    \\
Streptococcus agalactiae (ATCC BAA-611)   & GCF\_000007265.1   & 1.8                     & 7                    \\
Streptococcus mutans (ATCC 700610)        & GCF\_000007465.2   & 18                      & 5                   
\end{tabular}
\label{tab:atcc}
\end{table}

\begin{table}[ht]
\centering
\caption{Dataset Information - Phylotag/16S}
\begin{tabular}{lcccc}
\toprule
\textbf{Species}                & \textbf{Accession} & \textbf{Abundance (\%)} & \textbf{Copy Number} &  \\
\midrule
Acetivibrio thermocellus        & GCF\_000015865.1   & 0.3451                  & 4                    &  \\
Allomeiothermus silvanus        & GCF\_000092125.1   & 13.11167                & 2                    &  \\
Clostridium perfringens         & GCF\_000013285.1   & 0.14404                 & 8                    &  \\
Coraliomargarita akajimensis    & GCF\_000025905.1   & 5.09543                 & 2                    &  \\
Corynebacterium glutamicum      & GCF\_000011325.1   & 0.51815                 & 6                    &  \\
Desulfoscipio gibsoniae         & GCF\_000233715.2   & 4.83635                 & 8                    &  \\
Desulfosporosinus acidiphilus   & GCF\_000255115.2   & 12.08939                & 9                    &  \\
Desulfosporosinus meridiei      & GCF\_000231385.2   & 4.34622                 & 11                   &  \\
Echinicola vietnamensis         & GCF\_000325705.1   & 0.57616                 & 4                    &  \\
Escherichia coli                & GCF\_000005845.2   & 0.28808                 & 7                    &  \\
Fervidobacterium pennivorans    & GCF\_000235405.2   & 5.84364                 & 1                    &  \\
Frateuria aurantia              & GCF\_000242255.2   & 6.90793                 & 4                    &  \\
Hirschia baltica                & GCF\_000023785.1   & 8.54939                 & 2                    &  \\
Nocardiopsis dassonvillei       & GCF\_000092985.1   & 0.04301                 & 5                    &  \\
Olsenella uli                   & GCF\_000143845.1   & 1.89953                 & 1                    &  \\
Salmonella bongori              & GCF\_000252995.1   & 0.25907                 & 7                    &  \\
Salmonella enterica             & GCF\_000018625.1   & 0.80623                 & 7                    &  \\
Sediminispirochaeta smaragdinae & GCF\_000143985.1   & 10.3629                 & 2                    &  \\
Segniliparus rotundus           & GCF\_000092825.1   & 2.8498                  & 1                    &  \\
Streptococcus pyogenes          & GCF\_000006785.2   & 0.3451                  & 6                    &  \\
Stutzerimonas stutzeri          & GCF\_000327065.1   & 3.2239                  & 4                    &  \\
Terriglobus roseus              & GCF\_000265425.1   & 5.03741                 & 2                    &  \\
Thermobacillus composti         & GCF\_000227705.2   & 12.5215                 & 5                    & 
\end{tabular}
\label{tab:phylotag}
\end{table}

\fi

\end{document}